\documentclass[fleqn,12pt,twoside]{article}
\usepackage{amsmath}
\usepackage{espcrc1}

\usepackage{graphicx}
\usepackage[figuresright]{rotating}

\title{New mean field theory with the parity and charge mixing for the pion in nuclei}

\author{Kiyomi Ikeda\address[RIKEN]{Institute of Physical and
Chemical Research (RIKEN),\\ Wako, Saitama 351-0198, Japan},
        Satoru Sugimoto\addressmark[RIKEN],
    and
    Hiroshi Toki\address{Research Center for Nuclear Physics (RCNP),
    Osaka University,\\ Ibaraki, Osaka 567-0047,
    Japan}\addressmark[RIKEN]
}

\begin{document}
\maketitle
\begin{abstract}
We give a brief review on the important role of the pion-exchange
interaction  characterized by the strong tensor force.  In order
to illustrate the role of the pion, we have proposed a new  mean
field theory with the charge and parity mixing, by which the pion
is treated explicitly in the mean field framework.  We introduce
the essence of the new mean field theory and illustrate the
calculated results of the pion (tensor force) correlation  by
taking the alpha particle as an example.
\end{abstract}
\section{Introduction}
We propose a mean field theory to describe nuclei by treating the
pion explicitly.   In the non-relativistic framework, the
pion-exchange interaction appears mainly as the tensor force,
which is much stronger than the central force. To make the
important role of the pion in nuclei clear, we briefly discuss the
role of the tensor force in the reaction matrix theory, showing
the density dependence of the central component of the $G$-matrix,
together with some discussion on the results of the variational
calculations for light nuclei based on the realistic nuclear
force. The importance of the pion demonstrated through those
studies, leads us to develop a theoretical framework based on the
mean field theory which enables us to treat the pion-exchange
interaction on the same footing as other meson-exchange
interactions. A new mean field theory has been formulated by
breaking the charge and parity symmetries of the single-particle
orbits and further by recovering both the symmetries of the total
wave function. Thus we call it the charge- and parity-projected
Hartree-Fock (CPPHF) scheme, with which we study the correlation
caused by the pion (or tensor force).  We discuss the CPPHF scheme
in the latter half of this paper.
\section{Standard model  for nuclei}
The reaction matrix theory initiated by Br\"{u}ckner has been
developed with the intention of understanding the nuclear
structure through the multiple scattering correlation based on the
two-particle scattering in nuclei starting from the realistic
nuclear force. The most basic element in the theory is the
reaction matrix ($G$-matrix). The introduction of the $G$-matrix
is necessitated by the singular nature of the two-body forces in
the short range region, that is, the repulsive core and the strong
non-central components.
\begin{figure}[htb]
\begin{center}
\includegraphics[width=30mm]{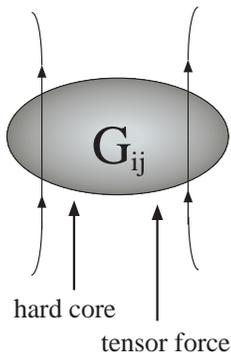}
\caption{The $G$-matrix for two particles in a nucleus, which
includes the effects of the hard core and the tensor force.}
\label{fig1}
\end{center}
\end{figure}

The elementary process of the multiple scattering is the
two-nucleon scattering within a nucleus which is described by
``the reaction matrix'' given by  $G=V+V\{Q/e\}G$ with
$e=\omega-QTQ$, where $V$, $T$, $\omega$, and $Q$ express the
two-boy nuclear potential, the kinetic energy operator, the
starting energy of the scattering two nucleons, and the Pauli
exclusion operator, respectively.   The reaction matrix $G$ is a
sort of ``effective interaction'' in nuclei in contrast to the
original nuclear force $V$.   Because the two-body scattering
correlation is known to be dominant compared with higher multiple
scattering correlations, the reaction matrix $G$ can be a good
basic element for investigating the nuclear structure with the
standard models (the shell model and the mean field model).

\section{Renormalization of  the tensor force  effect  in the
effective interactions} According to the definition of $G$, the
repulsive core of $V$ appears simply in $G$ as a fairly weakened
core. However, the tensor force causes the significant
modifications in the characteristics of the ``effective
interaction $G$,'' because the effect of the tensor force arises
first as the second order effect $V_\mathrm{T}\{Q/e\}V_\mathrm{T}$
and produces a very large contribution to the central component of
$G$. The many-body effects included in the $G$-matrix arises only
through the Green function $Q/e$. These are the Pauli effect due
to $Q$ and the dispersion effect due to $1/e$, which bring about
various kinds of the structure dependence reflecting the
characteristics of the structure assumed initially. Thus the
contribution of the tensor force to the central component of $G$
is very sensitive to the nuclear structure itself and to the
circumstance of the surrounding nuclear medium \cite{akaishi72}.
We discuss here the density dependence of $G$ as a typical example
of the structure dependence.

\begin{figure}[htb]
\begin{minipage}[t]{75mm}
\includegraphics[width=75mm]{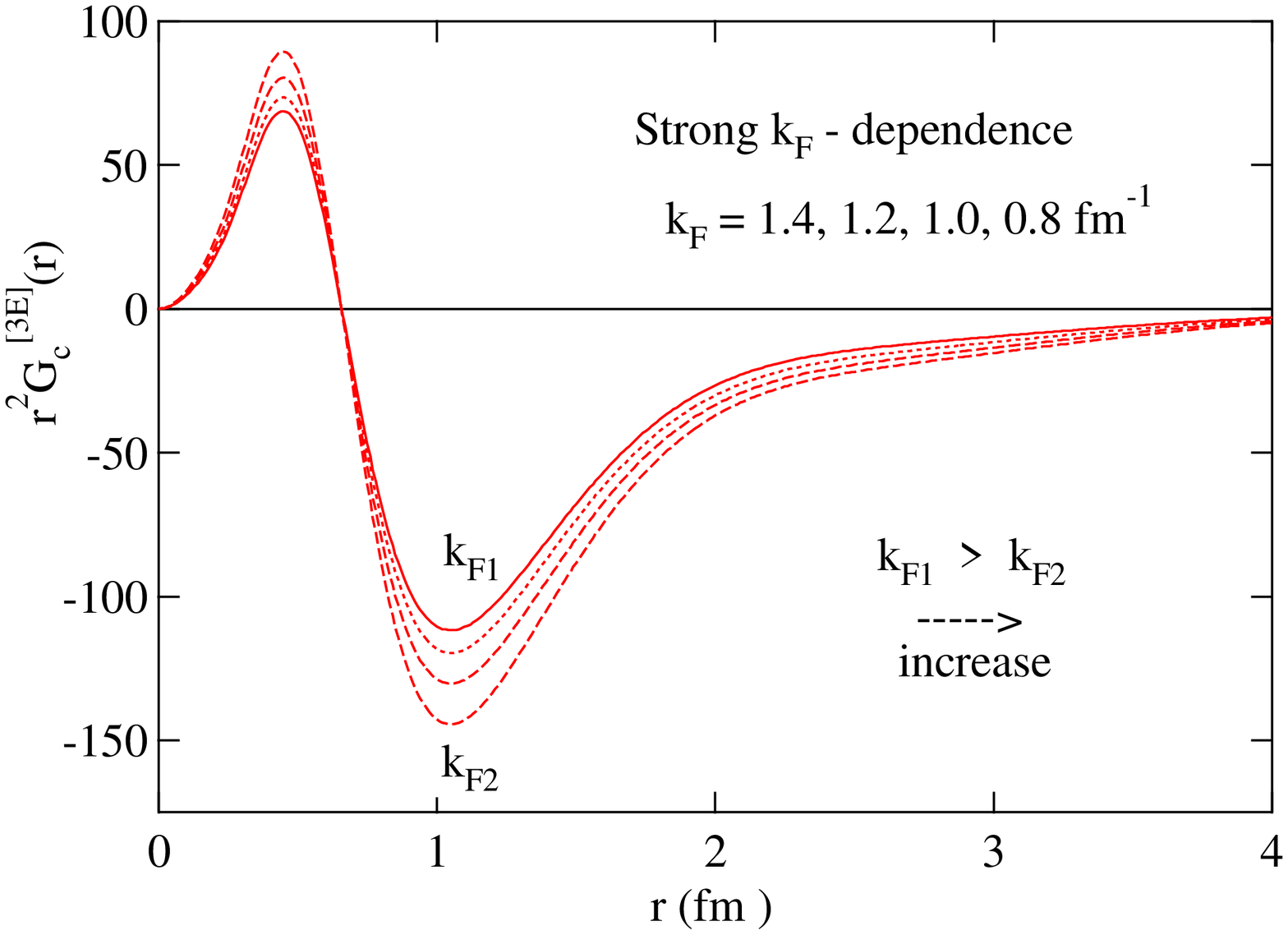}
\caption{$r^2G_\mathrm{C}^{[^3\mathrm{E}]}$ for
$k_\mathrm{F}$=1.4, 1.2, 1.0, and 0.8 fm${^{-1}}$, where the
tensor force $V_\mathrm{T}$ is included. The attraction in the
intermediate and long range ($r > 1$ fm) increases with decreasing
the Fermi momentum.\label{fig2}}
\end{minipage}
\hspace{\fill}
\begin{minipage}[t]{75mm}
\includegraphics[width=75mm]{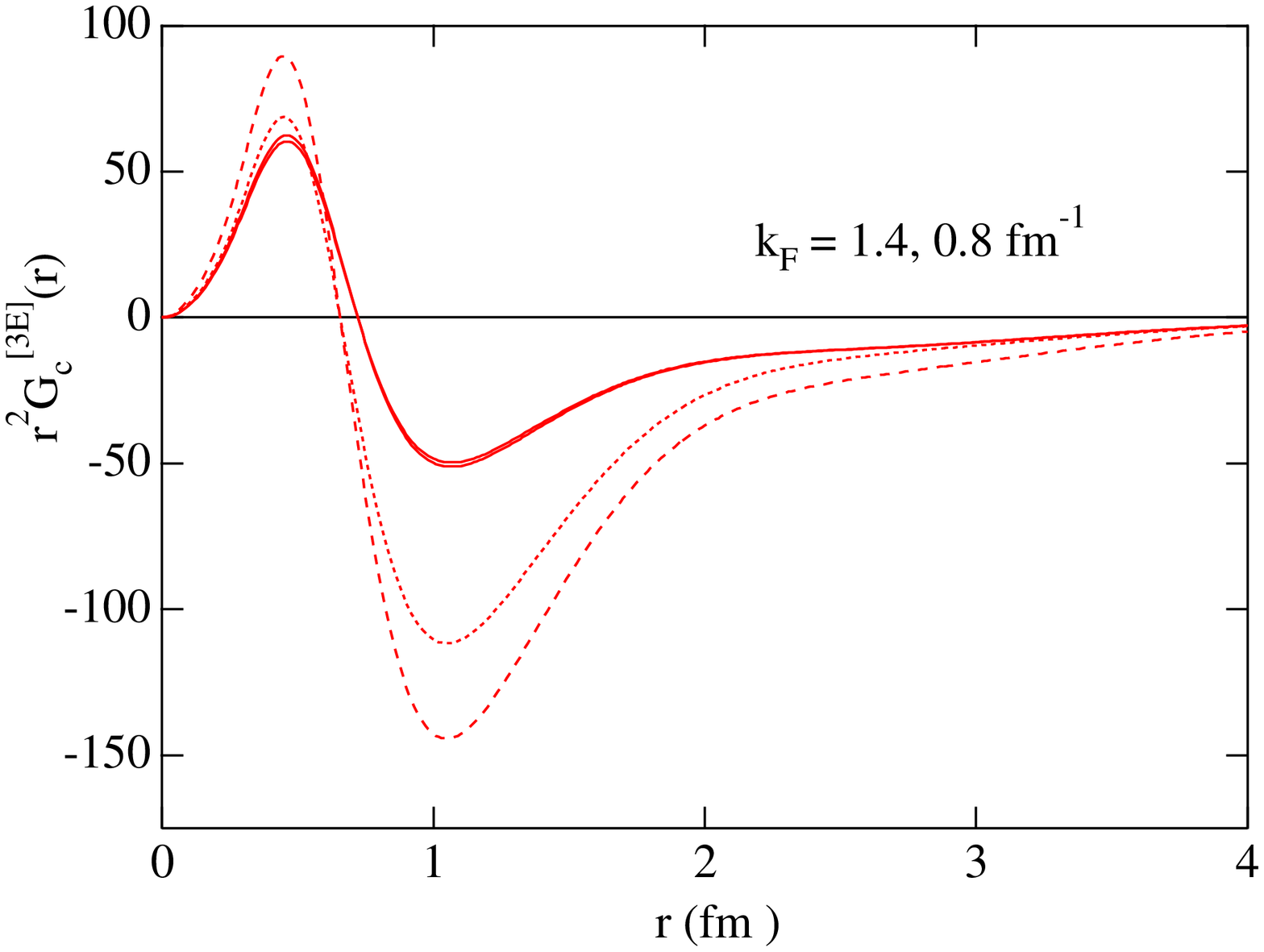} \caption{
The comparison between $r^2\tilde{G}_\mathrm{C}^{[^3\mathrm{E}]}$
(solid curves) and $r^2 G_\mathrm{C}^{[^3\mathrm{E}]}$ (dashed
curves) for $k_\mathrm{F}$=1.4 and 0.8 fm$^{-1}$. The difference
$(r^2 G_\mathrm{C}^{[^3\mathrm{E}]}-r^2
\tilde{G}_\mathrm{C}^{[^3\mathrm{E}]})$ corresponds to
$V_\mathrm{T}(Q/e)V_\mathrm{T}$.\label{fig3}}
\end{minipage}
\end{figure}
We show the central component of $G$ for the triplet-even state,
$G_\mathrm{C}^{[^3\mathrm{E}]}$ with the two-body interaction $V$
in which the tensor force $V_\mathrm{T}$ is included, and
$\tilde{G}_\mathrm{C}^{[^3\mathrm{E}]}$ with
$\tilde{V}=V-V_\mathrm{T}$ in which tensor force is not included,
which are calculated by Y.~Akaishi. Figure~\ref{fig2} clearly
shows the strong $k_\mathrm{F}$ dependence of
$G_\mathrm{C}^{[^3\mathrm{E}]}$. On the contrary, there is little
difference between two cases of
$\tilde{G}_\mathrm{C}^{[^3\mathrm{E}]}$ for $k_\mathrm{F}$=1.4 and
0.8 fm$^{-1}$ as shown in Fig.~\ref{fig3} by the solid curves. It
is easily understood from Fig.~\ref{fig3} that the density
($k_\mathrm{F}$) dependence of $G_\mathrm{C}^{[^3\mathrm{E}]}$ is
almost entirely caused by the tensor force and that the tensor
contribution $V_\mathrm{T}\{Q/e\}V_\mathrm{T}$ is estimated by the
difference
$(G_\mathrm{C}^{[^3\mathrm{E}]}-\tilde{G}_\mathrm{C}^{[^3\mathrm{E}]})$.
Figure~\ref{fig3} shows that the ratio
$(G_\mathrm{C}^{[^3\mathrm{E}]}-\tilde{G}_\mathrm{C}^{[^3\mathrm{E}]})/
\tilde{G}_\mathrm{C}^{[^3\mathrm{E}]}$ = 1.5 $\sim$ 2.0 at $r$=1.0
fm. These features tell us that the tensor force plays the very
important role on the nuclear structures.

\section{Variational calculation in the real space}
Recent variational calculations of light nuclei based on the
realistic nuclear force in the real space are carried out by the
Argonne-Illinois Group up to A=10 \cite{wiringa00}.   First of all
we have been strongly impressed by their successful reproduction
of the binding energy of the ground and excited states for all
these nuclei. Secondly, we are surprised by their results that the
contribution of the one-pion-exchange interaction $V^\pi$ amounts
to 70-80\% of the strength of the two-body interaction part in the
3$\le$A$\le$8 systems, and the pion is also important for the
three-body interaction originating mostly from the pion exchange.
Thirdly, we notice that the two-alpha cluster structure is found
in $^8$Be: According to their expression \cite{wiringa00}, ``these
results obtained from the VMC wave functions, suggest that the
$0^+$, $2^+$, $4^+$ wave functions for $^{8}$Be have the structure
of a deformed rotor consisting of two $\alpha$'s.''

With regard to the importance of the pion, we note here the
pioneering work of the variational calculation of the $\alpha$
particle by ATMS \cite{akaishi86}. Table~\ref{table1} exhibits the
contributions of various interactions to the binding energy (in
the unit of MeV), from which we can see almost a half of the
attraction is due to the tensor force and hence the pion-exchange
interaction is responsible in alpha particle.
\begin{table}[hbt]
\caption{The breakdown of energies for the $\alpha$ particle in
the ATMS calculations \cite{akaishi86}.\label{table1}}
\begin{tabular}[t]{cccccccccc}
\hline
 Energy&KE&PE& $^3$E$_\mathrm{C}$&
 $^1$E$_\mathrm{C}$&$^1$O$_\mathrm{C}$+$^3$O$_\mathrm{C}$&$^3$E$_\mathrm{T}$&$^3$O$_\mathrm{T}$&LS+QLS&P(D)\\
 \hline
 $-$20.6&131.1&$-$157.7&$-$51.3&$-$26.2&$-$0.4&$-$69.7&$-$0.5&$-$3.6&12.8\%\\
\hline
\end{tabular}
\end{table}

\section{New mean field theory with the parity and charge mixing}
In this section,  we construct a new framework of the mean field
model  under the assumption that if we introduce a suitably wide
space spanned by a set of single-particle basis, we can describe
the tensor correlation with fairly high-momentum components. We
also assumed that the short-range correlation caused by the
repulsive core can be treated, separately from the tensor
correlation, by the $G$-matrix theory with the wide model space
mentioned above. We assume the effective interaction in the
following form;
\begin{align}
G(r)=\tilde{G}(r)+\Omega (r) V_\mathrm{T} (r),
\end{align}
where $\tilde{G}$ is the $G$-matrix by adopting the two-body
interaction $\tilde{V}$(= $V-V_\mathrm{T}$ ) and $\Omega(r)$  is a
correlation function acting between $S$-wave and $D$-wave states
induced by the tensor force.

Let us consider that a virtual pion is emitted and absorbed by a
nucleon in a  good parity orbit. Then  the relevant nucleon makes
a jump from one single-particle state to another with the opposite
parity, accompanied by a spin-flip due to the pion-nucleon
coupling. Therefore, to incorporate the effect of the correlation
caused by the pion-exchange interaction in the parity-conserved
single-particle space, we must treat the higher configurations,
like 2p--2h (two-particle--two-hole) states, crossing over major
shells. If we want to treat the pion in the framework of a mean
field theory, the parity symmetry of single-particle orbits in
nuclei should be necessarily broken. Similarly, the emission and
absorption of a virtual pion with charge  require the change of
the charge of a nucleon single-particle orbit in nuclei.
\begin{figure}[hbt]
\begin{center}
\includegraphics[width=50mm]{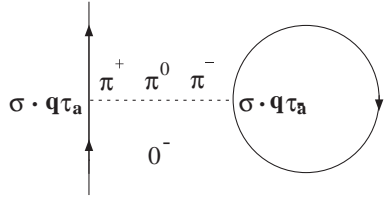}
\caption{The mean field diagram for the treatment of the tensor
force with three pion charge states.\label{fig7}}
\end{center}
\end{figure}

In the mean field theory with the pion, a single-particle orbit
consists of the four components with different charge-parity of
[p,$+$], [p,$-$], [n,$+$], and [n,$-$];
\begin{align}
\psi_{\alpha_i}(i)= \psi_{\alpha_i}^{[p,+]}(i)
+\psi_{\alpha_i}^{[p,-]}(i)+\psi_{\alpha_i}^{[n,+]}(i)
+\psi_{\alpha_i}^{[n,-]}(i),
\end{align}
The intrinsic wave function is defined by a Slater determinant of
a set \{$\psi_{\alpha_i}$\};
\begin{align}
\Psi^{\mathrm{intr}}=\frac{1}{\sqrt{A!}}\hat{\mathcal{A}}
\prod_{i=1}^{A}
 \psi_{\alpha_i}
 ,
\end{align}
which is a trial wave function with the parity and charge mixing.
Generally it is necessary to recover the parity and charge
symmetries of the total wave function since the  nuclear state has
a definite charge and parity.   By using the following projection
operators, we can obtain the total wave function with a good
charge ($Z$) and parity ($+$ or $-$).
\begin{align}
\Psi^{[Z,\pm]}&=\hat{\mathcal{P}}^\mathrm{c}(Z)
\hat{\mathcal{P}}^\mathrm{p}(\pm)\Psi^{\mathrm{intr}},\label{eq5}\\
\hat{\mathcal{P}}^\mathrm{c}(Z)&= \frac{1}{2\pi} \int_0^{2 \pi} d \theta e^{i(\hat{Z}-Z)\theta}
,\quad \hat{Z}=\sum_{i=1}^A\frac{1+\tau_i^3}{2},\\
\hat{\mathcal{P}}^\mathrm{p}(\pm)&=\frac{1\pm \hat{P}}{2},\quad
\hat{P}=\prod_{i=1}^A \hat{p}_i .
\end{align}

Here, we illustrate the characteristic of the parity-projected
wave functions for the doubly-closed-shell nuclei;
\begin{align}
\hat{\mathcal{P}}^\mathrm{p}(+)\Psi^{\mathrm{intr}}&=|[\text{0p--0h}]\rangle+|[\text{2p--2h}]\rangle+|[\text{4p--4h}]\rangle+\cdots,\\
\hat{\mathcal{P}}^\mathrm{p}(-)\Psi^{\mathrm{intr}}&=|[\text{1p--1h}]\rangle+|[\text{3p--3h}]\rangle+\cdots
.
\end{align}
The positive-parity state consists of even number of
$\text{1p--1h}$ pairs with $0^-$. This means that the
positive-parity projection provides $\text{2p--2h}$ states as the
major correction terms. Hence, the correlated wave function after
the parity projection provides the $\text{2p--2h}$ admixture due
to the pion-exchange interaction. This admixture corresponds to
the $D$-state admixture in the $\alpha$ particle. The $D$-state
probability in the $\alpha$ particle is known to be around
10--15\%. The negative-parity state has odd number of
$\text{1p--1h}$ pairs with $0^-$ spin-parity.

\begin{figure}[hbt]
\begin{center}
\includegraphics[width=80mm]{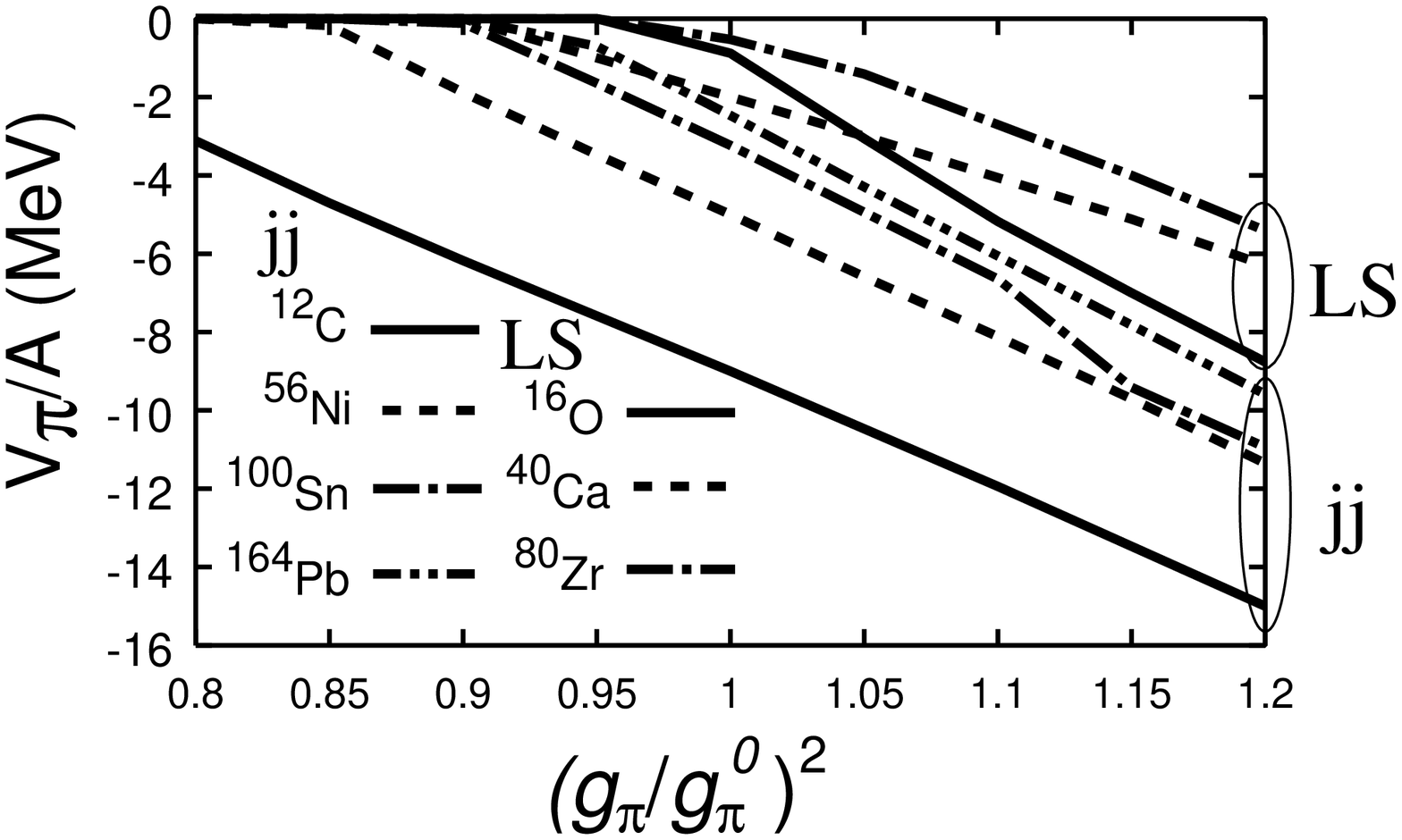}
\caption{The pion energy per nucleon as a function of the
pion-nucleon coupling constant square.  In this systematic study,
we have fixed the set of the filled single-particle states with
the same spins as in the closed-shell configurations.\label{fig9}}
\end{center}
\end{figure}
The first attempt of the study with the new mean field theory with
the pion was made in the framework of the relativistic mean field
theory by solving the equation obtained from the variation $\delta
\langle \Psi^\mathrm{intr}|H|\Psi^\mathrm{intr} \rangle=0$, where
the Hamiltonian $H$ is obtained from the Lagrangian including the
pion terms \cite{toki02}. We obtain the solution of a parity-mixed
self-consistent Hartree field with the finite pion mean field
($\langle\pi\rangle\neq 0$) by taking the free-space pion-nucleon
coupling constant ($g^0_{\pi})$.

In  Fig.~\ref{fig9}, the calculated pion energy  per nucleon  for
various nuclei with the closed-shell configuration is plotted as a
function of the pion-nucleon coupling constant ($g_\pi$). The
definition of the critical coupling constant $g_\pi^\mathrm{cr}$
is one at which there arises a finite pion mean field for each
nucleus. The values of $g_\pi^\mathrm{cr}$ for various nuclei
except for $^{12}$C are distributed in the region satisfying
\begin{align}
0.92 g_{\pi}^0 \le g_\pi^\mathrm{cr} \le 0.97 g_{\pi}^0.
\end{align}
This fact indicates that ``the parity breaking mean field is
fragile'' in the sense that various effects not taken into account
in the present study could influence the realization of the finite
pion mean field.

Second attempt of our study in the new mean field theory is the
charge- and parity-projected Hartree-Fock study with the tensor
force of light nuclei.  An essential  improvement made in this
study is that we adopt  the charge- and parity-projected wave
function giving in Eq.~(\ref{eq5}) as a variational wave function
and then we intend to solve the finite mean field based on the
variation after the charge and parity projections:
\begin{align}
\delta
\frac{\langle\Psi^{[Z,\pm]}|H|\Psi^{[Z,\pm]}\rangle}{\langle\Psi^{[Z,\pm]}|\Psi^{[Z,\pm]}\rangle}=0
.
\end{align}
The variation with respect to each single-particle orbit leads to
the charge- and parity-projected Hartree-Fock (CPPHF) equation.
When we compare the CPPHF equation with the ordinary Hartree-Fock
(HF) equation, the CPPHF equation has a much more complicated form
but its structure remains similar to the HF equation.

\begin{table}[hbt]
\caption{Results for the ground ($0^+$) state of the alpha
particle in
 various cases.
The potential energy ($\langle \hat{v} \rangle$ in MeV), the
kinetic energy ($\langle \hat{T} \rangle$ in MeV), the total
energy ($E$ in MeV), the root-mean-square matter radius
($R_\mathrm{m}$ in fm) and the probability of the $p$-state
component (P(-)) are given. $\langle \hat{v}_\mathrm{C} \rangle$,
$\langle \hat{v}_\mathrm{T} \rangle$, $\langle \hat{v}_\mathrm{LS}
\rangle$, and $\langle \hat{v}_\mathrm{Coul} \rangle$ are the
expectation values for the central, the tensor, the LS, and the
Coulomb potentials, respectively (in MeV). \label{table2}}
\begin{center}
\begin{tabular}[t]{crrr}
\hline & \multicolumn{1}{c}{HF}&
\multicolumn{1}{c}{PPHF}&\multicolumn{1}{c}{CPPHF}\\
\hline
$\langle \hat{v}_\mathrm{C} \rangle$&$-$56.85&  $-$61.31& $-$64.75\\
$\langle \hat{v}_\mathrm{T} \rangle$&0.00&  $-$10.91&$-$30.59\\
$\langle \hat{v}_\mathrm{LS} \rangle$&0.00&    0.67&1.91\\
$\langle \hat{v}_\mathrm{Coul} \rangle$&0.76&    0.78&0.85\\
$\langle \hat{v} \rangle$&        $-$56.10&  $-$70.76&$-$92.58\\
$\langle \hat{T} \rangle$&       39.98&   49.67&64.39\\
$E$&                             $-$16.12&  $-$21.09&$-$28.19\\\
$R_\mathrm{m}$&                   1.63&    1.50&1.37\\
P(-)&                             0.00&    0.08& 0.16\\
\hline
\end{tabular}
\end{center}
\end{table}
We solve the CPPHF equation for the $\alpha$ particle, which is
the most simple doubly-closed-shell nucleus. Here, the calculated
results for the ground ($0^+$) state are exhibited in
Table~\ref{table2} for the purpose to examine the responsibility
of the CPPHF scheme to describe the tensor correlation. We compare
them with the results for the cases of the Hartree-Fock (HF)
scheme with the non-projected variational wave function and for
the case of the parity-projected Hartree-Fock (PPHF) scheme with
the parity-projected variational wave function. Although we make
the tensor force 1.5 times stronger than the usual one, we cannot
obtain the correlation energy due to the tensor force by the HF
scheme but a fairly large correlation energy is obtained by the
PPHF scheme. Finally the CPPHF scheme produces a large amount of
the correlation energy, which is almost three times larger than
that for the PPHF case. (See the detail results and discussions in
Ref. \cite{sugimoto04}.)

\section{Conclusion}
The studies in the model space based on the $G$-matrix theory and
in the realistic
     functional space based on the realistic nuclear force, exhibit clearly
the importance of
     the tensor force (the  one-pion-exchange force).
     A new framework is introduced to treat the pionic correlation in
the extended
      model space by introducing the single-particle orbits with the parity
and charge
      mixing and by recovering the parity and charge symmetries by the
      projection.
We consider that the new theoretical framework will provide a
powerful method to investigate the mechanism of forming diverse
structures of nuclei (shell structure, cluster structure, halo
structure and others), which are explored by the experiments.
\section*{Acknowledgements}
We acknowledge fruitful discussions with Prof. Y.~Akaishi and
Dr.~Y.~Ogawa on the role of the pion in light nuclei.

\end{document}